\newcommand{\tstern}{\ensuremath{T_2^*} }
\def \deg {\ensuremath{^\circ} }

\newcommand*{\fg}[1]{Fig.\thinspace\ref{#1}}
\newcommand*{\fgs}[1]{Figs.\thinspace\ref{#1}}
\newcommand*{\eq}[1]{Eq.\thinspace\ref{#1}}

\documentclass[aps,prl,preprint,superscriptaddress]{revtex4}

\usepackage{graphicx}
\usepackage{dcolumn}
\usepackage{bm}
\usepackage[english]{babel}

\begin{document}

\preprint{V1 -- \today{}}

\title{Two-dimensional optical control of electron spin orientation by
linearly polarized light in InGaAs}

\author{K. Schmalbuch}
\author{S. G\"{o}bbels}
\author{Ph. Sch\"{a}fers}
\author{Ch. Rodenb\"{u}cher}
\author{P. Schlammes}
\affiliation{II. Institute of Physics, RWTH Aachen University,
Otto-Blumenthal-Stra{\ss}e, 52074 Aachen, Germany}
\affiliation{JARA: Fundamentals of Future Information Technology,
J\"{u}lich-Aachen Research Alliance, Aachen, Germany}

\author{Th. Sch\"{a}pers}
\author{M. Lepsa}
\affiliation{Institute of Bio- and Nanosystems, Forschungszentrum
J\"{u}lich, 52428 J\"{u}lich, Germany} \affiliation{JARA:
Fundamentals of Future Information Technology, J\"{u}lich-Aachen
Research Alliance, Aachen, Germany}

\author{G. G\"{u}ntherodt}
\author{B. Beschoten}
\email{beschoten@physik.rwth-aachen.de} \affiliation{II. Institute
of Physics, RWTH Aachen University, Otto-Blumenthal-Stra{\ss}e,
52074 Aachen, Germany} \affiliation{JARA: Fundamentals of Future
Information Technology, J\"{u}lich-Aachen Research Alliance, Aachen,
Germany}

\date{\today}

\begin{abstract}

Optical absorption of circularly polarized light is well known to
yield an electron spin polarization in direct band gap
semiconductors. We demonstrate that electron spins can even be
generated with high efficiency by absorption of linearly polarized
light in In$_x$Ga$_{1-x}$As. By changing the incident linear
polarization direction we can selectively excite spins both in polar
and transverse directions. These directions can be identified by the
phase during spin precession using time-resolved Faraday rotation.
We show that the spin orientations do not depend on the crystal axes
suggesting an extrinsic excitation mechanism.

\end{abstract}

\maketitle

The generation of spin-polarized charge carriers by optical
orientation in non-magnetic semiconductors is well-established. In
optical orientation the angular momentum of circularly polarized
photons will be transferred to electrons and holes during absorption
\cite{Chapter02_Dyakonov-Perel_TheoryofOpticalSpinOrientationofElectronsandNucleiinSemiconductors,
Zutic_RMP_2004}. This can result in a large spin polarization of
$50\%$ in bulk III-V semiconductors. Besides static imaging and
probing of the spin polarization
\cite{PRL94_Crooker2005_ImagingSpinFlowsinSemiconductorsSubjecttoElectricMagneticandStrainFields,PRB66_Dzhioev2002_Low-temperatureSpinRelaxationinN-typeGaAs},
optical pump probe measurements using time-resolved Faraday rotation
(TRFR) has become a standard method both for triggering and probing
of spin coherence in semiconductors
\cite{PRL80_Kikkawa1998_ResonantSpinAmplificationinN-TypeGaAs,
Kikkawa_Nature_1999, Salis_Nature_2001,
PRB63_Kimel2001_Room-TemperatureUltrafastCarrierandSpinDynamicsinGaAsProbedbythePhotoinducedMagneto-opticalKerrEffect,
Kato_Nature_2004,
Science313_Greilich2006_ModeLockingofElectronSpinCoherencesinSinglyChargedQuantumDots,
NaturePhysics3_Meier2007_MeasurementofRashbaandDresselhausSpin-OrbitMagneticFields,
PRB75_Schreiber2007_AnisotropicElectronSpinLifetimein(InGa)As-GaAs(110)QuantumWells}.
In contrast, optical absorption of linearly polarized photons should
not result in net spin polarization as an equal number of spin-up
and spin-down electrons and holes will be generated. Yet it is known
that illumination of III-V-semiconductor quantum wells with linearly
polarized light can yield a spin-dependent photo-voltage response
originating from spin photo-galvanic effects
\cite{Ganichev_nat_phys_2006,
SST23_Bel'kov2008_Magneto-GyrotropicEffectsinSemiconductorQuantumWells}.
However, those spin currents have not been probed by magneto-optical
methods.

In this Letter we report on measurements of electron spin coherence
in InGaAs by TRFR after optical orientation by linearly polarized
laser pulses. We observe both polar and transverse initial spin
orientations, which can be controlled independently as a function of
the incident linear polarization direction. The number of transverse
spins increases linearly with the magnitude of a perpendicular
external magnetic field and vanishes at zero field, while the number
of polar spins is unaffected by the external magnetic field. We
demonstrate that the optical orientation by linearly polarized light
has a comparable efficiency as that by circularly polarized light.
The generated respective spin orientations are furthermore found to
be independent of the crystal axes orientation of the InGaAs layer
indicating an extrinsic excitation mechanism.

We have studied several $\mathrm{In}_x\mathrm{Ga}_{1-x}\mathrm{As}$
samples with In-contents $0\leq x\leq0.1$ and thicknesses between
300~nm and 1~$\mu$m, which are grown by molecular beam epitaxy on
semi-insulating (001)GaAs substrates. The room temperature carrier
density was set to $n \sim 3 \times 10^{16}$ cm$^{-3}$ by Si
co-doping to allow for long spin dephasing times at low temperatures
\cite{PRL80_Kikkawa1998_ResonantSpinAmplificationinN-TypeGaAs,PRB66_Dzhioev2002_Low-temperatureSpinRelaxationinN-typeGaAs}.
Note that all presented results have been observed in all samples
independent of In content and thickness. In the following, we show
representative data, which were taken on a 500~nm epilayer of
$\mathrm{In}_x\mathrm{Ga}_{1-x}\mathrm{As}$ with $x=4.9~\%$.
\begin{figure}[bp]
\includegraphics{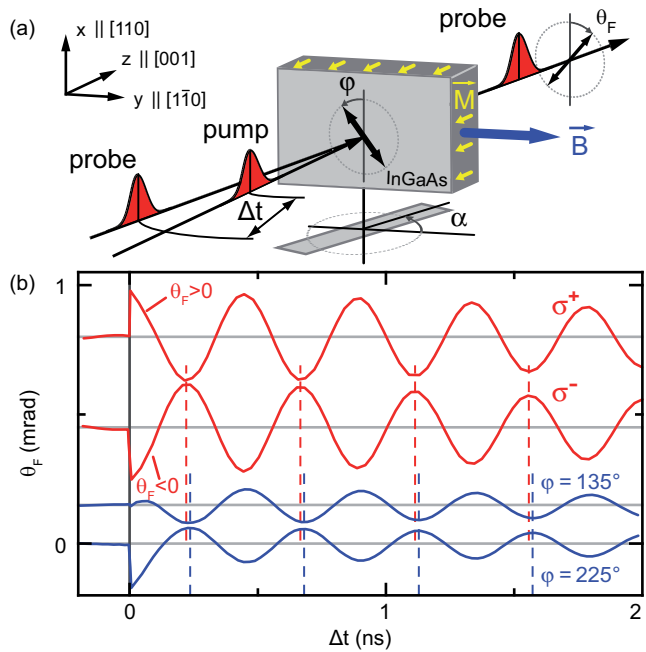}
\caption{\label{fig1} (Color). (a)~Setup for all-optical measurements
of TRFR using linearly polarized pump pulses. The incident linear
polarization direction can continuously be adjusted by the angle
$\varphi$. (b)~Comparison of TRFR after excitation by circularly
(red) and linearly (blue) polarized pump pulses. Data are taken at
$T=30$~K in a transverse magnetic field of 0.5~T. The dashed
vertical lines indicate a phase shift in the precession between
circular and linear excitation. A vertical offset is added for
clarity. }
\end{figure}
The sample was mounted strain-free in an optical He-flow cryostat.
Phase triggering of electron spin coherence is achieved either by
circularly or linearly polarized ps pump pulses. The incident
polarization direction of the latter can be changed continuously by
an angle $\varphi$ as defined in \fg{fig1}a. Spin precession is
probed in a transverse external magnetic field $B$ (oriented along
the $y$-direction) by a second time-delayed linearly polarized probe
pulse using standard measurements of the TRFR angle $\theta_F$,
which is a measure of the polar spin component. Its time dependent
evolution can be described by an exponentially damped cosine
function
\begin{equation}
\theta_F(\Delta t)=\theta_0\cdot\exp\left(-\frac{\Delta
t}{\tstern}\right)\cdot\cos\left(\omega_L \Delta t+\delta\right),
\label{Fitformel}
\end{equation}
with amplitude $\theta_0$, transverse spin dephasing time $\tstern$,
Larmor frequency $\omega_L=g \mu _B B/\hbar$, time delay $\Delta t$
between the pump and probe beam and phase factor $\delta$. $g$ is
the effective electron $g$-factor, $\mu_B$ the Bohr magneton, and
$\hbar$ the Planck constant. The sample plane can furthermore be
rotated by an angle $\alpha$ about the $x$-axis.

\fg{fig1}b depicts TRFR measurements after optical excitation with
both circularly and linearly polarized pump pulses using a laser
energy near the fundamental band gap of the InGaAs layer. Data were
taken at $T=30$~K and $B=0.5$~T. When using circularly polarized
light the angular momentum of the photon will be absorbed during
interband absorption resulting in spin-polarized electrons and
holes. The holes will be ignored in the following discussion. By
changing the light helicity from $\sigma^+$ to $\sigma^-$, we can
control the initial spin orientation between parallel and
antiparallel alignment relative to the incident light propagation
direction. The resulting TRFR measurements are depicted in
\fg{fig1}b (red curves) for $\sigma^+ /\sigma^-$ laser excitation
under nearly normal incidence. The change of the initial spin
orientation is easily seen by a sign change of $\theta_F$ right
after excitation ($\Delta t=0$~ns). Note that we measure $\theta_F$
in the polar geometry for all presented experiments. We are thus
only sensitive to spin components, which are pointing in the $\pm
z$-direction.
\begin{figure}[tp]
\includegraphics{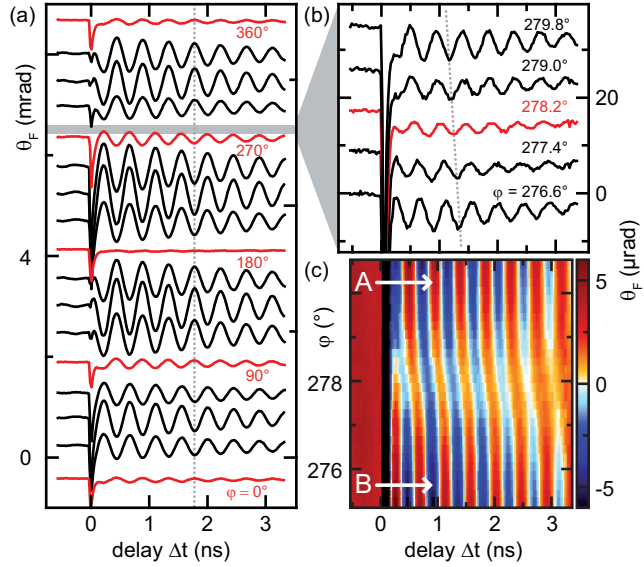}
\caption{\label{fig2} (Color). TRFR for optical excitation with
linearly polarized laser pulses measured in $n$-InGaAs at $T=30$~K
and $B=0.5$~T for (a) $\varphi =0...360\deg$ (the vertical dotted
line helps to see the sign reversal every $90\deg$). The sign
reversal is accompanied by a continuous phase shift of the
precessing spins, which can be seen in~(b) and~(c). }
\end{figure}
When using linearly polarized pump pulses, we would not expect to
excite a net spin polarization as the linearly polarized light is a
superposition of $\sigma^+$ and $\sigma^-$ photons, which results
into an equal number of spin-up and spin-down electrons after
absorption. In \fg{fig1}b we show TRFR data taken for normal
incidence $\alpha =0\deg$ at two distinct polarization angles of
linearly polarized pump pulses, which differ by $90\deg$ (blue
curves). We, however, clearly observe spin precession. Surprisingly,
the amplitude is only slightly reduced compared to the curves taken
under $\sigma^+ /\sigma^-$ excitation demonstrating that optical
orientation of electron spins by linearly polarized light is
strikingly efficient.

We note that spins of opposite directions can be excited when
changing the polarization by $90\deg$ from $135\deg$ to $225\deg$. While
the precession frequency is identical for all excitations, we
observe a phase shift in the precession for linear polarized
excitation as indicated by the vertical dashed lines in $\sigma^+
/\sigma^-$ in \fg{fig1}b. Such a phase shift indicates a change of
the initial spin direction.

To further explore the polarization dependence of $\theta_F$, we
plot a series of TRFR measurements with $\varphi$ varying between
$0\deg$ and $360\deg$ in \fg{fig2}a. The most striking observation
is the sign reversal of $\theta_F$ every $90\deg$ (see also
\fg{fig4}a), which is further investigated in \fgs{fig2}b and~c,
where the polarization angle resolution is enhanced in a regime of
sign reversal. It is clearly seen that spin precession is observed
at all angles. The sign reversal of $\theta_F$ between point~A
($\varphi =280.2\deg$, red color code $\theta_F>0\deg$) and point~B
($\varphi =275.8\deg$, blue color code $\theta_F<0\deg$) in Fig.~2c
is accompanied by a continuous change of the phase $\delta$ of the
precessing spins (see also dotted line in \fg{fig2}b).
\begin{figure}[tbp]
\includegraphics{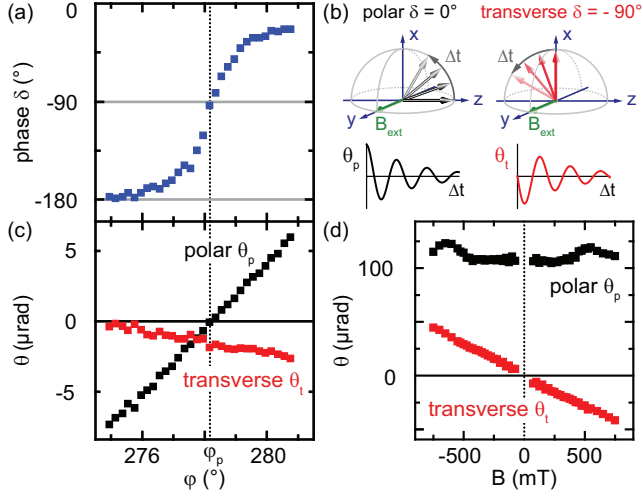}
\caption{\label{fig3} (Color). (a) Phase of precessing spins after
optical excitation with linearly polarized light at different
incident polarization angles $\varphi$ near the sign reversal in
\fg{fig2}b,~c. (b)~Illustration of spins oriented in polar and
transverse direction and respective TRFR curve as expected in the
polar observation direction. According to the measured phase,
$\theta_F$ will be decomposed into polar and transverse amplitudes
plotted as a function of (c)~incident linear polarization direction
and (d)~external magnetic field.}
\end{figure}
The phase can be extracted from fitting all TRFR traces in
\fg{fig2}c by \eq{Fitformel}. As seen in \fg{fig3}a, the phase
continuously changes from $-180\deg$ to almost $0\deg$ within a
small range of linear polarization angles. We want to emphasize that
in our polar configuration we only probe spins, which have a finite
projection along the $\pm z$-axis. For $\delta=0\deg$, spins are
oriented in the polar $+z$ direction at $\Delta t=0$, which results
in a TRFR curve starting in a positive maximum (see also
\fg{fig3}b). On the other hand, spins are oriented in the $-z$
direction for $\delta =-180\deg$. This explains the sign reversal
between points~A and~B in \fg{fig2}c. In contrast, for a phase of
$\delta =-90\deg$ spin precession starts with $\theta_F=0\deg$ at
$\Delta t=0$~ns, which is illustrated in \fg{fig3}b. This
unambiguously demonstrates that spins will be oriented along the
$x$-axis (transverse to both incident light direction and magnetic
field direction) at the respective linear polarization angle. It is
important to note that the excitation of transverse spins is unique
to the optical orientation with linearly polarized light. It has not
been observed for excitation with circularly polarized light.
Knowing that we can excite both polar and transverse spins with
linearly polarized light, we can now decompose the projections for
any polarization angle along the polar and the transverse direction
using $\theta_{\textit{p}}=\theta_0\cdot \cos{\delta}$ and
$\theta_{\textit{t}}=\theta_0\cdot \sin{\delta}$, respectively. As
shown in \fg{fig3}c, polar and transverse spin signals change sign
at different pump polarization angles. By carefully adjusting the
polarization angle $\varphi$ we achieve a full 2-dimensional control
over the initial spin orientation, which cannot be realized by any
other electrical or optical technique.

\begin{figure}[tbp]
\includegraphics{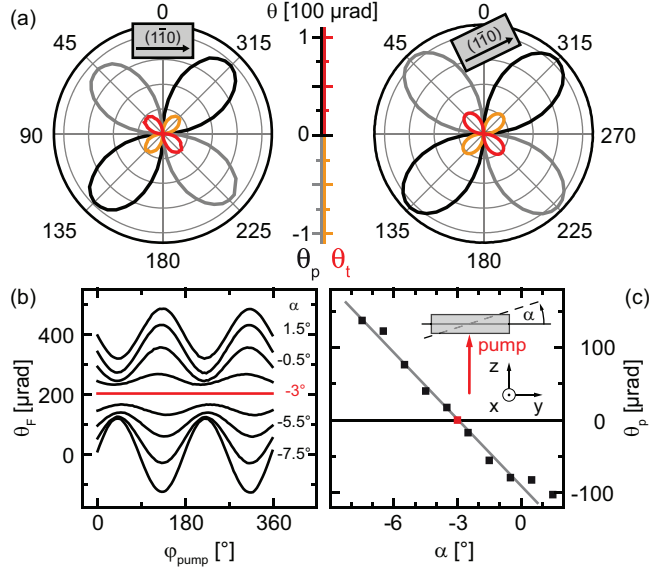}
\caption{\label{fig4} (Color). (a)~Polar diagrams of $\theta_p$
(black/grey) and $\theta_t$ (red/orange) as a function of incident
polarization direction for two different crystal orientations.
(b)~$\theta_p$ vs. $\varphi$ at various angles of incidence
$\alpha$. A vertical offset is added for clarity. (c)~Linear
dependence of $\theta_p$ plotted at $\varphi =40\deg$. The sign
reversal at $\alpha =-3\deg$ shows a change in spin orientation of
the polar spins near normal incidence.}
\end{figure}

To further explore the origin of these unexpected findings, we plot
the magnetic field dependence of both polar and transverse spin
components in \fg{fig3}d. Note that the amplitude is a measure of
the number of electron spins initially oriented in the respective
direction. The amplitude of the polar spins (black squares) is
almost independent of magnetic field, while the amplitude of the
transverse spins (red squares) depends linearly on the magnetic
field and vanishes at $B=0$~T. This shows that transverse spins can
only be excited at non-zero magnetic fields. We note that the
transverse spin orientation switches sign from $+x$ to $-x$
direction for negative $B$ values. Such a linear $B$ dependence has
previously been reported in photocurrent measurements where spins
are optically generated by infrared absorption. The underlying
intrinsic theory of the so-called magnetogyrotropic photogalvanic
effect
\cite{SST23_Bel'kov2008_Magneto-GyrotropicEffectsinSemiconductorQuantumWells}
relates the spin orientation directly to the underlying crystal
structure. We can easily test whether the observed effect is related
to the crystal axis by comparing the linear polarization dependence
of $\theta_0$ for different sample orientations. In \fg{fig4}a we
compare the dependence of the Faraday amplitudes
$\theta_{\textit{p}}$ and $\theta_{\textit{t}}$ on $\varphi$ with
the $[1\bar{1}0]$ crystal axis either oriented parallel (left panel)
or under $26\deg$ (right panel) with respect to the external
magnetic field direction (see also \fg{fig1}a) using polar plots.
The amplitudes have been extracted from the TRFR data using the
above analysis \cite{remark}. It is obvious that the observed
symmetry is independent of the crystal orientation. The same
behavior was also observed for other angles at 45 and $90\deg$ (not
shown) suggesting that our effect is of extrinsic origin. This
excludes various intrinsic effects as a source for the observed spin
polarization, such as spin polarization induced by the Dresselhaus
fields
\cite{PR100_Dresselhaus1955_Spin-OrbitCouplingEffectsinZincBlendeStructures},
intrinsic spin Hall effect
\cite{PRL92_Sinova2004_UniversalIntrinsicSpinHallEffect}, intrinsic
double refraction and dichroism
\cite{PLA174_Bungay1993_SpecularOpticalActivityinGaAs,CPL217_Bungay1994_TimeNon-InvariantLinearBirefringenceandDichroismduetoSpin--OrbitInteraction},
inverse Faraday and Cotton-Mouton effect
\cite{Nature435_Kimel2005_UltrafastNon-ThermalControlofMagnetizationbyInstantaneousPhotomagneticPulses,PRL99_Kalashnikova2007_ImpulsiveGenerationofCoherentMagnonsbyLinearlyPolarizedLightintheEasy-PlaneAntiferromagnetFeBO3}
or piezo-optical effects
\cite{SSC97_Koopmans1996_AQuantitativeStudyofthePiezoopticalActivityinGaAs}.

We note that the observed spin polarization becomes largest for both
polar and transverse spin components at an incident polarization
direction of $45\deg$ (see \fg{fig4}a), while it vanishes for $s$
and $p$ polarized light at $\varphi=0\deg$ and $90\deg$,
respectively. Away from normal incidence, it is well-known, that the
linear polarization state is conserved only for $s$ and $p$
polarized light. For other polarization angles, there will be an
additional birefringence, which becomes largest for
$\varphi=45\deg$. At this angle, the light is elliptically
polarized, which would result in spin orientation by the circular
polarization component along the laser propagation direction. As the
birefringence is negligible near normal incidence, it will be
instructive to study the spin polarization away from normal
incidence. We therefore rotated the sample about the $x$-axis by an
angle $\alpha$ as defined in \fg{fig1}a. In \fgs{fig4}b and~c, we
focus on the polar spin amplitude $\theta_p$. The data in \fg{fig4}b
were taken at a fixed pump-probe delay of $\Delta t=1$~ns with
$B=0$~T for different values of $\alpha$. As expected, the sign of
$\theta_p$ oscillates as a function of $\varphi$. Surprisingly,
there is a sign reversal of $\theta_p$ at $\alpha =-3\deg$, which is
not seen for excitation with circularly polarized light (not shown),
showing that the polar spin orientation can be switched into
opposite orientations when rotating the sample from $\alpha <-3\deg$
to $\alpha >-3\deg$ (Fig.~4c). This furthermore explains why we
observe spins at normal incidence, i.e. at $\alpha =0\deg$. Note
that there is a remarkable increase of $\theta_p$ slightly away from
$\alpha=-3\deg$. At $\alpha = <-7.5\deg$ we find a value of
$\theta_p$, which is $85\%$ of the corresponding amplitude obtained
for $\sigma^+$ excitation. Such a large number of spins cannot be
explained by the above birefringence effect, which is estimated to
be on the order of a few percent at $\alpha = -7.5\deg$. We should
emphasize that the birefringence in our optical setup is less than
$1\%$ and thus can also not explain the observed spin polarization.

Another possible source of spin polarization are local electric
fields induced by the laser pulse. It was shown previously that
static electric fields can indeed induce a spin polarization as
demonstrated, e.g., in experiments on current-induced spin
polarization
\cite{PRL93_Kato2004_Current-InducedSpinPolarizationinStrainedSemiconductors}
or extrinsic spin Hall effect
\cite{Science306_Kato2004_ObservationoftheSpinHallEffectinSemiconductors}.
In our ultrafast optical experiments, the ps laser pulse generates a
non-equilibrium electron-hole population. As the mobility of holes
is much lower than for electrons, there will be a fast built-up of
local electric fields. In fact, Nuss \emph{et al.}
\cite{PRL58_Nuss1987_DirectSubpicosecondMeasurementofCarrierMobilityofPhotoexcitedElectronsinGalliumArsenide}
showed that electron diffusion in GaAs starts on timescales much
shorter than a ps. Although we currently do not understand the
complex linear light polarization dependence of the spin population,
it is, however, likely that these local electric fields resulting
from the non-equilibrium carrier distribution are relevant for the
observed effect.

In conclusion, we have demonstrated that optical orientation of
electron spins by linearly polarized light is very effective. From
the phase during spin precession we can unambiguously prove that
spins can be generated in both polar and transverse directions. This
is in contrast to optical orientation by circularly polarized light
for which spins are aligned in polar directions only. Surprisingly,
the spin orientation is independent of the in-plane crystallographic
directions of the sample, pointing to an extrinsic origin of the
spin polarization. The full two-dimensional control over the initial
spin direction offers to study anisotropies in spin relaxation. This
might give decisive clues about the dominating spin generation and
relaxation mechanisms.

This work was supported by DFG through FOR 912.
\bibliographystyle{apsrev}

\selectlanguage{english}


\begin{thebibliography}{26}
\bibitem[{rem()}]{remark}
  \bibinfo{title}{There is a slight amplitude difference for polarization
  rotations of $\Delta\varphi=180\deg$, which are related the retardation plate
  and which has been subtracted for simplicity.}
\end{thebibliography}

\end{document}